\documentclass[a4paper,11pt]{article}

\usepackage{fullpage}
\usepackage[T1]{fontenc}

\usepackage{amssymb,amsmath,cite}
\usepackage{hyperref,comment,xcolor,bbold,booktabs}

\usepackage{amsthm}

\begin{document}
\vspace{5mm}
\vspace{0.5cm}

\def\thefootnote{\arabic{footnote}}
\setcounter{footnote}{0}

\allowdisplaybreaks 

\begin{titlepage}
\thispagestyle{empty}

\begin{flushright}
	\hfill{\ MPP-2022-26} 
\end{flushright}
				
\vspace{35pt}
				
\begin{center}
	{\LARGE{\bf Weak gravity versus scale separation}}
									
	\vspace{50pt}
							
	{\Large Niccol\`o Cribiori$^{1}$ and Gianguido Dall'Agata$^{2,3}$}
							
	\vspace{25pt}
							
	{
	{\it $^{1}$Max-Planck-Institut f\"ur Physik (Werner-Heisenberg-Institut),\\ 
	F\"ohringer Ring 6, 80805 M\"unchen, Germany}
										
		\vspace{15pt}

		{\it  $^{2}$Dipartimento di Fisica e Astronomia ``Galileo Galilei''\\
			Universit\`a di Padova, Via Marzolo 8, 35131 Padova, Italy}
										
		\vspace{15pt}
										
		{\it   $^{3}$INFN, Sezione di Padova \\
		Via Marzolo 8, 35131 Padova, Italy}
		}
								
\vspace{40pt}
								
{ABSTRACT} 
\end{center}

We give evidence that fully supersymmetric Anti-de Sitter vacua of extended supergravity with a residual gauge group containing an abelian factor cannot be scale separated as a consequence of the weak gravity conjecture.
We prove this for $N=2$ and $N=8$ supergravity and we explain how our argument applies also to vacua with partially broken, but extended residual supersymmetry. We finally discuss possible loopholes and especially how certain $N=1$ models can evade our reasoning. Our results suggest that $N=0,1$ supersymmetry at the lagrangian level might be the most promising chances to obtain a truly four-dimensional effective description of quantum gravity.

\vspace{10pt}
			
\bigskip
			
\end{titlepage}

\numberwithin{equation}{section}

\baselineskip 6 mm

\section{Introduction} 

From the point of view of a low energy observer our universe looks four-dimensional. 
However, the existence of extra dimensions is not ruled out experimentally and it is actually required by string theory for consistency.
One has thus to explain why the typical size of the four observed spacetime dimensions is much larger than that of the yet unobserved additional ones. 
The puzzling existence of this hierarchy of scales goes commonly under the name of scale separation problem.

Aiming at investigating realistic scenarios, one should address such a problem within setups with positive background energy density, such as de Sitter vacua or quintessence. 
This would however introduce further complications, related to the unavoidable breaking of supersymmetry. 
In this work, we thus take a modest attitude and concentrate on the scale separation problem for effective theories with Anti-de Sitter vacua and a certain amount of preserved supercharges. 

Low energy effective theories coupled to quantum gravity should obey consistency criteria known as swampland conjectures (see e.g.~\cite{Palti:2019pca,vanBeest:2021lhn} for reviews). 
Among these, some have consequences for the fate of scale separation. More specifically, the Anti-de Sitter distance conjecture \cite{Lust:2019zwm} implies that the size of the Anti-de Sitter spacetime and that of the internal manifold cannot be parametrically decoupled. 
Further refinements have been proposed in \cite{Blumenhagen:2019vgj}, introducing logarithmic corrections, and in \cite{Buratti:2020kda}, considering the effect of discrete symmetries. 
These statements are supported by several examples already discussed in the corresponding articles, (see also \cite{Lust:2020npd} for a recent analysis of two-dimensional Anti-de Sitter solutions). 
On the other hand, certain classes of string theory compactifications \cite{DeWolfe:2005uu}, based on earlier works \cite{Behrndt:2004km,Derendinger:2004jn,Lust:2004ig}, are known to be in tension with these conjectures.\footnote{A systematic search for scale separation in large classes of four-dimensional anti-de Sitter vacua from a ten-dimensional perspective has been initiated in \cite{Tsimpis:2012tu}. See instead \cite{Farakos:2020phe}, for three-dimensional scale-separated Anti-de Sitter flux vacua recently constructed in type IIA. See also \cite{Apers:2022zjx}, for more comments on various properties of these three- and four-dimensional Anti-de Sitter type IIA vacua in relation to swampland conjectures.} 
Such models have been revisited recently and their understanding has been improved beyond the smeared approximation for the involved sources \cite{Junghans:2020acz, Marchesano:2020qvg, Cribiori:2021djm,Emelin:2022cac}. 
At present, it is not clear if something is missing in the analysis in such a way that the Anti-de Sitter distance conjecture is eventually restored. The scale separation problem is thus still an open, fundamental question even for Anti-de Sitter vacua. 

While this is clearly far from a phenomenologically acceptable setup, we should note that most of the scenarios that try to generate a positive cosmological constant in string theory start from an Anti-de Sitter vacuum and then add small, typically non-perturbative, effects to uplift the vacuum energy \cite{Kachru:2003aw,Balasubramanian:2005zx}.
Our analysis is therefore certainly relevant also for this situation.

In this work, we do not assume any of the aforementioned swampland criteria directly against scale separation. 
Instead, we show that the absence of scale separation follows as a consequence of yet another conjecture, the (magnetic) weak gravity conjecture \cite{Arkani-Hamed:2006emk}, which is among the oldest and arguably most tested ones.\footnote{Notably, an argument connecting weak gravity to scale separation appeared already in the original paper \cite{Arkani-Hamed:2006emk}.} 
In the spirit of the swampland program, we will employ a bottom-up perspective and require that the magnetic weak gravity conjecture applies to any effective theory, while remaining agnostic on the possible string theory origin. 
We will consider four-dimensional extended supergravity theories admitting supersymmetric Anti-de Sitter vacua with an unbroken gauge group containing an abelian factor. 
For these models, we will show that the cosmological constant is of the order of the ultraviolet cutoff dictated by the weak gravity conjecture, namely the gauge coupling
\begin{equation}
\label{Hquant}
|\mathcal{V}_{AdS}| \gtrsim {q}^2 \, g^2 M_P^2 \gtrsim {q}^2\, \Lambda_{UV}^2.
\end{equation}
We will additionally assume the charge ${q}$ to be non-vanishing and quantized, but the precise details of charge quantisation will not be important. 
What matters for us is that in the expression of the vacuum energy there is no parameter which can be (parametrically) sent to zero. 
Taking the Kaluza--Klein scale as a proxy for the ultraviolet cutoff,  $\Lambda_{KK} \sim \Lambda_{UV}$, the absence of (parametrical) scale separation for vacua such that \eqref{Hquant} follows, because
\begin{equation}
\frac{|\mathcal{V}_{AdS}| }{\Lambda^2_{KK}} \gtrsim 1 .
\end{equation}
We will give an explicit proof that $N=2$ and $N=8$ Anti-de Sitter vacua with generic gaugings do obey \eqref{Hquant}. 
We will explain how our argument should apply also to $N$-extended partial supersymmetry breaking vacua of $N'>N$ supergravity theories, explicitly showing that this is the case for partially broken $N=2$ vacua of $N=8$ supergravity.
Finally, we discuss how $N=0,1$ models, such as \cite{DeWolfe:2005uu}, can evade our argument if they do not arise as consistent truncations of a higher supersymmetric theory.

Our methodology is analogous to the one employed in \cite{Cribiori:2020use, DallAgata:2021nnr} to exclude de Sitter vacua in extended supergravity. This suggests that difficulties in constructing de Sitter vacua are somehow related to difficulties in obtain scale separation (see \cite{Gautason:2015tig} for a similar reasoning from a different perspective).

This work is organised as follows. 
In section \ref{sec:_n_2_supergravity}, we present our general argument in $N=2$ supergravity and we apply it to a concrete example of Anti-de Sitter vacua arising from M-theory compactified on Sasaki-Einstein manifolds. In section \ref{sec:_n_8_supergravity}, we repeat the argument in $N=8$ supergravity, which is the maximal theory in four-dimensions. In section \ref{sec:partial_susy}, we discuss generic patterns of partially broken $N=2$ vacua of $N=8$ supergravity and show that the argument applies also in those cases. This gives support to the fact that it should indeed apply to general Anti-de Sitter vacua with extended supersymmetry, or at least to those obtainable as partial breaking within a higher supersymmetric theory. Finally, we discuss possible loopholes and emphasise that $N=0,1$ vacua seem to be the most promising candidates to reach scale separation. We work in Planck units.

\section{$N=2$ supergravity} 
\label{sec:_n_2_supergravity}

We start by proving that four-dimensional $N=2$ Anti-de Sitter vacua with an unbroken gauge group containing an abelian factor are of the form \eqref{Hquant}. 
Through this section, we  follow the conventions of \cite{Ceresole:1995ca,Andrianopoli:1996cm}.

\subsection{The argument}
\label{N=2arg}

The vacuum energy of fully supersymmetric Anti-de Sitter vacua is given by the trace of the gravitino mass matrix
\begin{equation}
\mathcal{V}_{AdS} = - 3  \bar L^\Lambda  L^\Sigma \mathcal{P}^x_\Lambda \mathcal{P}^x_\Sigma,
\label{VN2AdS}
\end{equation}
where $L^\Lambda$ are the symplectic sections of the special K\"ahler manifold and $\mathcal{P}^x_\Lambda$ are the quaternionic moment maps. 
Fully supersymmetric solutions of four-dimensional $N=2$ gauged supergravity have been discussed in detail in \cite{Hristov:2009uj}. 
Among the vacuum conditions given there, we recall that
\begin{equation}
\mathcal{P}^x_\Lambda f^\Lambda_i =0,
\end{equation}
where $f^\Lambda_i \equiv D_i L^\Lambda$ is the K\"ahler covariant derivative of the symplectic sections. 
Squaring and using the special geometry identity $g^{i \bar \jmath} f_i^\Lambda f_{\bar\jmath}^\Sigma =  -\frac12 \left({\rm Im} \mathcal{N}^{-1}\right)^{\Lambda \Sigma} - \bar L^\Sigma L^\Lambda$, we get
\begin{equation}
\label{ImNm32}
 \bar L^\Lambda  L^\Sigma \left(\mathcal{P}^x_\Lambda \mathcal{P}^x_\Sigma +\mathcal{P}^0_\Lambda \mathcal{P}^0_\Sigma\right) = -\frac12 \left({\rm Im} \mathcal{N}^{-1}\right)^{\Lambda \Sigma} \mathcal{P}_\Lambda^x \mathcal{P}_\Sigma^x ,
\end{equation}
where $g_{i\bar\jmath}$ is the K\"ahler metric, $({\rm Im} \mathcal{N})_{\Lambda \Sigma}$ the kinetic matrix of the vector fields and $\mathcal{P}^0_\Lambda$ the special K\"ahler moment map. 
This relation is central in our argument, because it expresses the gravitino mass in terms of the gauge couplings. 
Notice that we also used the special geometry relation $\left({\rm Im} \mathcal{N}^{-1}\right)^{\Lambda \Sigma} \mathcal{P}_\Lambda^0 \mathcal{P}_\Sigma^0 =-2 g_{i \bar\jmath} k^i_\Lambda k^{\bar\jmath}_\Sigma \bar L^\Lambda L^\Sigma$ \cite{Ceresole:1995ca,Andrianopoli:2001zh}, where $k^i_\Lambda$ are special K\"ahler Killing vectors, and notice that this quantity vanishes on the vacuum \cite{Hristov:2009uj}. 
Thus, we can express the vacuum energy entirely in terms of the moment maps, containing the gauge charges and couplings
\begin{equation}
\mathcal{V}_{AdS} =\frac32 \ ({\rm Im}\, \mathcal{N}^{-1})^{\Lambda \Sigma} \left(\mathcal{P}^0_\Lambda \mathcal{P}^0_\Sigma+\mathcal{P}^x_\Lambda \mathcal{P}^x_\Sigma\right).
\end{equation}
Since we want to show that this expression is related to the square of the charge of the gravitinos on the vacuum, it is convenient to define an SU(2) charge matrix
\begin{equation}
\label{QSU2}
2 {Q_{\Lambda\, A}}^B = \mathcal{P}_\Lambda^0 \delta_A^B + {(\sigma_x)_A}^B \mathcal{P}_\Lambda^x,
\end{equation}
such that the gravitino covariant derivative takes the form $D_\mu \psi_{\nu \, A} = \dots + i A_\mu^\Lambda  {Q_{\Lambda\, A}}^B  \psi_{\nu\, B}$
and the vacuum energy can be rewritten as
\begin{equation}
\label{VN2AdSTrQQ}
 \mathcal{V}_{AdS} = 3  \ ({\rm Im}\, \mathcal{N}^{-1})^{\Lambda \Sigma}\,  \,{\rm Tr}\, Q_\Lambda Q_\Sigma,
\end{equation}
where
\begin{equation}
{\rm Tr}\, Q_\Lambda Q_\Sigma\equiv {Q_{\Lambda\, A}}^B {Q_{\Sigma\, B}}^A =   \frac12\left(\mathcal{P}^0_\Lambda \mathcal{P}^0_\Sigma+\mathcal{P}^x_\Lambda \mathcal{P}^x_\Sigma\right).
\end{equation}

To recast \eqref{VN2AdSTrQQ} into the form \eqref{Hquant}, we have first to identify the abelian vector associated to the weak gravity conjecture and canonically normalise its kinetic term. 
To this purpose, we split the vector fields as $A_\mu^\Lambda  =A^{\perp\, \Lambda}_\mu +A^{\parallel\, \Lambda}_\mu$ with
\begin{equation}
\label{Adec}
A^{\parallel\, \Lambda}_\mu = {{P^\parallel}^\Lambda}_\Sigma A^\Sigma_\mu \equiv \frac{({\rm Im} \mathcal{N}^{-1})^{\Lambda \Sigma}\Theta_\Sigma}{\Theta^2} \tilde A_\mu, \qquad  A^{\perp\, \Lambda}_\mu \equiv {{P^\perp}^\Lambda}_\Sigma A^\Sigma_\mu,
\end{equation}
where the projectors ${{P^\parallel}^\Lambda}_\Sigma$, ${{P^\perp}^\Lambda}_\Sigma$, defined as in \cite{Cribiori:2020use} by
\begin{equation}
		{{P^\parallel}^\Lambda}_\Sigma = \frac{({\rm Im} \mathcal{N}^{-1})^{\Lambda \Gamma}\Theta_\Gamma \Theta_{\Sigma}}{\Theta^2} , \qquad {{P^\perp}^\Lambda}_\Sigma  = \delta^\Lambda{}_{\Sigma} - {{P^\parallel}^\Lambda}_\Sigma, 
\end{equation}
split the vectors into the combination contributing to the U(1) factor and the orthogonal ones. 
The vector gauging the weak gravity abelian factor is chosen to be
\begin{equation}
\label{Atilde}
\tilde A_\mu = \Theta_\Lambda A^\Lambda_\mu,
\end{equation}
for some coefficients $\Theta_\Lambda$. 
Substituting these expressions into the vectors' kinetic term, we get
\begin{equation}
\frac14 ({\rm Im} \mathcal{N})_{\Lambda \Sigma} F^\Lambda_{\mu\nu} F^{\Sigma\, \mu\nu} = \frac14 ({\rm Im} \mathcal{N})_{\Lambda \Sigma} F^\Lambda_{\mu\nu} (A^\perp) F^{\Sigma\, \mu\nu} (A^\perp) + \frac14 \frac{1}{\Theta^2} F_{\mu\nu} (\tilde A) F^{\mu\nu} (\tilde A),
\end{equation}
from which we identify the weak gravity gauge coupling \cite{Cribiori:2021gbf}
\begin{equation}
\label{g32U12}
g = \sqrt{-\Theta^2} = \sqrt{- \Theta_\Lambda ({\rm Im} \mathcal{N}^{-1})^{\Lambda \Sigma}\Theta_\Sigma}\,\, .
\end{equation} 
The kinetic term being canonically normalised, we have now to identify the charge. 
To this purpose, we repeat a similar procedure as above. 
We split the charge matrix \eqref{QSU2} as ${Q_{\Lambda}} = {Q^\parallel_{\Lambda}} + {Q^\perp_{\Lambda}}$ (SU(2) indices are understood when this is not source of confusion). 
Using also \eqref{Adec}, the gravitino covariant derivative becomes
\begin{equation}
\begin{aligned}
D_\mu \psi_{\nu \, A} =\dots i  A^{\perp\, \Lambda}_\mu {Q^\perp_{\Lambda\, A}}^B \psi_{\nu \, B} + i \tilde A_\mu {q_{A}}^B \psi_{\nu \, B},
\end{aligned}
\end{equation}
where the U(1) charge matrix is
\begin{equation}
{q_{A}}^B  =  \frac{\left({\rm Im}\mathcal{N}^{-1}\right)^{\Lambda \Sigma}\Theta_\Sigma}{\Theta^2} {Q^\parallel_{\Lambda\, A}}^B,
\end{equation}
as one can check by using $\left({\rm Im}\mathcal{N}^{-1}\right)^{\Lambda \Sigma}\Theta_\Sigma\,{Q^\perp_{\Lambda}} =0$.
From \eqref{Atilde} and $\tilde A_\mu {q}  =  A^{ \Lambda}_\mu {Q^\parallel_{\Lambda}}$, we find now that 
\begin{equation}
\label{Qthetaq}
 {Q^\parallel_{\Lambda\, A}}^B  = \Theta_\Lambda {q_A}^B,
\end{equation}
implying
\begin{equation}
{\rm Tr} Q_\Lambda^\parallel Q_\Sigma^\parallel = \Theta_\Lambda \Theta_\Sigma {\rm Tr}\, (q^2).
\end{equation}

Finally, we can look at the vacuum energy \eqref{VN2AdSTrQQ}, conveniently rewritten as
\begin{equation}
 \mathcal{V}_{AdS} = 3  \ ({\rm Im}\, \mathcal{N}^{-1})^{\Lambda \Sigma}  {\rm Tr} Q_\Lambda Q_\Sigma  =  3  \ ({\rm Im}\, \mathcal{N}^{-1})^{\Lambda \Sigma} \left( {\rm Tr} Q^\parallel_\Lambda Q^\parallel_\Sigma+{\rm Tr} Q^\perp_\Lambda Q^\perp_\Sigma\right),
\end{equation}
where the mixed term ${\rm Tr} Q^\parallel_\Lambda Q^\perp_\Sigma$ vanishes due to the projectors' orthogonality. 
Since the kinetic matrix $ ({\rm Im}\, \mathcal{N}^{-1})^{\Lambda \Sigma}  $ is negative definite, we have
\begin{equation}
 \mathcal{V}_{AdS}  \leq 3   \ ({\rm Im}\, \mathcal{N}^{-1})^{\Lambda \Sigma}  {\rm Tr} Q^\parallel_\Lambda Q^\parallel_\Sigma.
\end{equation} 
Inserting \eqref{Qthetaq} into this expression, we get eventually
\begin{equation}
\label{VAdSN=22}
\mathcal{V}_{AdS} \leq 3   \ ({\rm Im}\, \mathcal{N}^{-1})^{\Lambda \Sigma}  \Theta_\Lambda \Theta_\Sigma {\rm Tr} \, (q^2) = - 3 \, g^2\,  {\rm Tr} \, (q^2).
\end{equation}
In absolute value, this means
\begin{equation}
| \mathcal{V}_{AdS}  | \geq 3 \,g^2 \,  {\rm Tr} \, (q^2)  \gtrsim  \, {\rm Tr} \, (q^2)\,  \Lambda_{UV}^2,
\end{equation}
where in the last step we enforced the magnetic weak gravity conjecture. 
These vacua are therefore of the form \eqref{Hquant} and cannot be scale separated. 
We stress that our proof applies to any $N=2$ Anti-de Sitter vacuum with a generic (non-abelian) gauge group. 
The only working assumption we need is the presence of an unbroken abelian factor in the vacuum, which is required to apply the weak gravity conjecture.

While the argument above requires an abelian gauge group on the vacuum, we can argue that also a non-abelian gauge group would lead to the same conclusion, in a similar fashion to \cite{DallAgata:2021nnr}.
In fact, gauged extended supergravities contain the would-be goldstones of the non-abelian gauge symmetries and one could apply the argument above by looking at points in field space that are infinitesimallly close to the vacuum, but where the gauge group is broken to a subgroup containing an abelian factor.
In this case (which can be achieved in various ways, according to \cite{DallAgata:2021nnr}), the cut-off of the theories defined infinitesimally close to the vacuum must approach the cut-off of the non-abelian theory, while the U(1) gauge coupling will approach the non-abelian coupling.
Hence the application of the magnetic weak gravity conjecture for the theories close to the vacuum will give the desired results as a limit.

\subsection{An example: M-theory on Sasaki--Einstein manifolds}

To give a better understanding of how the above argument works in practice, we now provide a simple example.

Compactifications of M-theory on seven-dimensional Sasaki--Einstein manifolds give rise to N=2 supergravity with abelian gauging \cite{Gauntlett:2009zw, Hristov:2009uj}.
Following \cite{Hristov:2009uj}, we take the prepotential
\begin{equation}
F(X) =  \sqrt{X^0 (X^1)^3}
\end{equation}
and the symplectic sections to be
\begin{equation}
X^\Lambda = \left(\begin{array}{c}
1 \\[2mm] \tau^2
\end{array}\right), \qquad 
F_\Lambda = \left(\begin{array}{c}
\frac12 \tau^2
\\[2mm]
 \frac32 \tau
\end{array}\right).
\end{equation}
The K\"ahler potential is then
\begin{equation}
K = - \log \frac i2 (\tau- \bar \tau)^3.
\end{equation}
There is no gauging on the special K\"ahler manifold, thus $k^i_\Lambda = 0 = \mathcal{P}^0_\Lambda$.

The metric on the quaternionic manifold is
\begin{equation}
ds^2 = \frac{1}{4\rho^2} d \rho^2 + \frac{1}{4\rho^2}(d \sigma - i (\xi d \bar \xi-\bar \xi d \xi))^2 + \frac 1\rho d \xi d \bar \xi,
\end{equation}
where $\rho$ and $\sigma$ are real scalars, while $\xi$ is complex. We gauge the isometries generated by the abelian killing vectors 
\begin{equation}
\vec{k}_0 = 24 \partial_\sigma - 4i (\xi \partial_\xi - \bar \xi \partial_{\bar \xi}), \qquad \vec{k}_1 = 24 \partial_\sigma.
\end{equation}
The associated quaternionic prepotentials are
\begin{align}
\mathcal{P}_0^x &= \left(-4\sqrt \rho(\xi + \bar \xi), \,\, 4 i \sqrt \rho (\xi-\bar \xi),\,\, -\frac{12}{\rho} + 4\left(1- \frac{\xi \bar \xi}{\rho}\right)\right),\\
\mathcal{P}_1^x &= \left(0,0,-\frac{12}{\rho}\right).
\end{align}

The theory admits a supersymmetric Anti-de Sitter vacuum at
\begin{equation}
\xi =0, \qquad \tau = i, \qquad \rho=4,
\end{equation}
with vacuum energy 
\begin{equation}
\label{ccmodel}
 \mathcal{V}_{AdS} = -12.
\end{equation}
The field $\sigma$ does not appear in the scalar potential and is thus a free parameter. The gauge kinetic matrix of the vectors is diagonal on the vacuum
\begin{equation}
{\rm Im} \mathcal{N}_{\Lambda \Sigma} = \left( 
\begin{array}{cc}
 -\frac 12 & 0 \\
0 & -\frac 32\\
\end{array}
\right).
\end{equation}

The two killing vectors are gauging a U(1)$\times$U(1) isometry of the quaternionic manifold, but on the vacuum the prepotentials reduce to
\begin{equation}
\mathcal{P}_\Lambda^x = e_\Lambda \delta^{x3}, \qquad e_\Lambda = (1,-3),
\end{equation}
Therefore, the vacuum contains an unbroken U(1) factor. In the language of our general argument, we can here identify
\begin{equation}
2{{\mathcal{Q}^\parallel_\Lambda}_A}^B = {(\sigma_3)_A}^B e_\Lambda \equiv  2\Theta_\Lambda {q_A}^B,
\end{equation}
i.e.
\begin{equation}
 {q_A}^B= {(\sigma_3)_A}^B \mathrm{q}, \qquad \Theta_\Lambda = \frac{e_\Lambda}{2 \mathrm{q}}, \qquad {\rm Tr}\, (q^2) = 2 \mathrm{q}^2,
\end{equation}
where $\mathrm{q}$ is the abelian charge. The gauge coupling is thus
\begin{equation}
g^2 \mathrm{q}^2 = 2.
\end{equation}
Eventually, we see that the vacuum energy \eqref{ccmodel} agrees with \eqref{VAdSN=22} and it is then of the general form \eqref{Hquant}.

\section{$N=8$ supergravity}
\label{sec:_n_8_supergravity}

The structure of the argument presented in section \ref{N=2arg} is general, though the details depend on the specific $N=2$ structures.
To show that it can be easily extended to any gauged supergravity theory with $N>1$ supersymmetries, we will now show how it works for $N=8$ supergravity, which is the maximal theory in four dimensions. 
We follow the conventions of \cite{deWit:2007kvg} and, assuming that one might be less familiar with $N=8$ than with $N=2$ supergravity, we recall few facts about the former.

The couplings of the $N=8$ Lagrangian relevant for our analysis are the vectors' kinetic term, the scalar potential and the gravitino covariant derivative. 
The kinetic term is
\begin{equation}
e^{-1}\mathcal{L}_{kin} = \frac14 {\rm Im}\mathcal{N}_{\Lambda \Sigma} F^\Lambda_{\mu\nu}F^{\mu\nu\Sigma}, \qquad \Lambda, \Sigma = 0, \dots 27,
\end{equation}
where the kinetic matrix is negative definite and given by
\begin{equation}
\label{ImN=LL}
{({\rm Im}\mathcal{N}^{-1})}^{\Lambda \Sigma} = -2 L^{\Lambda}_{ij}L^{\Sigma ij}, \qquad i, j = 1,\dots 8,
\end{equation}
$L^{\Lambda}_{ij}$ being the coset representatives of the scalar manifold $E_{7(7)}/SU(8)$. 
The scalar potential is
\begin{equation}
\begin{aligned}
\mathcal{V} &= g^2 \left(\frac{1}{24}|{A_{2i}}^{jkl}|^2 - \frac34 |A_1^{ij}|^2\right)\\
&=\frac{g^2}{336} \mathcal{M}^{MN} \left(8 {\mathcal{P}_M}^{ijkl}\mathcal{P}_{Nijkl} + 9 {\mathcal{Q}_{Mi}}^j {\mathcal{Q}_{Nj}}^i\right),
\end{aligned}
\end{equation}
and we recall the relations \cite{deWit:2007kvg}\footnote{The indices $M,N=1,\dots, 56$ label symplectic vectors $V^M = (V^\Lambda, V_\Lambda)$ and are raised with the matrix  $\Omega^{MN} = \left(\begin{array}{cc}0 &1\\-1&0 \end{array}\right).$}
\begin{align}
\label{M=LL}
&\mathcal{M}_{MN} = {L_M}^{ij} L_{N\,ij} + L_{M\,ij} {L_N}^{ij},\\
&\mathcal{M}^{MN} {\mathcal{P}_M}^{ijkl}\mathcal{P}_{Nijkl}  = 4 |{A_{2i}}^{jkl}|^2,\\
\label{Qm32}
&\mathcal{M}^{MN}{\mathcal{Q}_{Mi}}^j{\mathcal{Q}_{Nj}}^i = -2  |{A_{2i}}^{jkl}|^2 - 28 |{A_1}^{ij}|^2.
\end{align}
The quantities ${A_1}^{ij}$ (gravitino mass) and ${A_{2i}}^{jkl}$ will also enter the supersymmetry transformations given below. 
The tensor $\mathcal{P}_{M\,ijkl}$ is defined as
\begin{equation}
\mathcal{P}_{M\,ijkl} = \frac{1}{24}\epsilon_{ijklmnpq}\mathcal{P}_M^{mnpq} = i \Omega^{NP} L_{N\,ij}{X_{MP}}^Q L_{Q\,kl},
\end{equation}
where ${X_{MN}}^{P}$ are the structure constants of the {\bf 56} representation of $E_{7(7)}$. 
The gauge charges ${\mathcal{Q}_{Mi}}^j$ are anti-hermitean SU(8) matrices, $\mathcal{Q}_M^\dagger = - \mathcal{Q}_M$, such that\footnote{We recall that complex conjugation exchanges the position of the SU(8) indices $i,j=1,\dots,8$.
In particular $({\mathcal{Q}_i}^j)^* = {\mathcal{Q}^j}_i = -{\mathcal{Q}_i{}^j}$ and therefore ${\mathcal{Q}_i}^j{\mathcal{Q}_j}^i =-{\mathcal{Q}^j}_i{\mathcal{Q}_j}^i= - {\rm Tr} \mathcal{Q}^\dagger Q$.}
\begin{equation}
{\mathcal{Q}_{M\,ij}}^{kl} = {\delta_{[i}}^{[k} {\mathcal{Q}_{M\,j]}}^{l]} = i \Omega^{NP}L_{N\,ij} {X_{MP}}^Q {L_Q}^{kl}.
\end{equation}
Finally, the gravitino covariant derivative contains the gauge connection term
\begin{equation}
\mathcal{D}_{\mu} \psi_{\nu\, i} = \dots - \frac{1}2 {A_\mu}^M {\mathcal{Q}_{M\,i}}^j\psi_{\nu j}.
\end{equation}
Since we want to study supersymmetric vacua, the supersymmetry transformations of the fermions are also relevant. 
Their form, up to three fermion terms, is
\begin{align}
&\delta \psi_\mu^i = \mathcal{D}_\mu \epsilon^i + \frac{\sqrt 2}{4}{\mathcal{H}}_{\rho\sigma}^{-\,ij} \gamma^{\rho\sigma}\gamma_\mu \epsilon_j + \sqrt 2  {A_1}^{ij} \gamma_\mu \epsilon_j,\\
&\delta \chi^{ijk} = -2 \sqrt 2 {\mathcal{P}_\mu}^{ijkl} \gamma^\mu \epsilon_l + \frac32 {\mathcal{H}}_{\rho\sigma}^{-\,[ij}\gamma^{\mu\nu}\epsilon^{k]} - 2 {A_{2l}}^{ijk}\epsilon^l.
\end{align}

\subsection{The argument}

We have first to determine the form of the scalar potential in a maximally supersymmetric Anti-de Sitter vacuum. 
Given that ${\mathcal{H}}_{\rho\sigma}^{\pm\,ij}$ is the (anti-)self dual part of the improved vector field strength ${\mathcal{H}}_{\rho\sigma}^{ij}$ (formula (2.51) of \cite{deWit:2007kvg}) and that $\mathcal{P}_{\mu\,ijkl} = i \Omega^{NP} L_{M\,ij}\mathcal{D}_\mu L_{N\,kl}$ contains derivatives of the scalars and a term with the gauge connection ${A_{\mu}}^M$, on a maximally symmetric vacuum we set
\begin{equation}
\label{maxsymmvac}
\mathcal{H}_{\mu\nu} =0= \mathcal{P}_{\mu}^{ijkl}.
\end{equation}
Furthermore, a maximally supersymmetric vacuum is such that all supersymmetry variations of the fermions vanish. 
Taking into account \eqref{maxsymmvac}, the condition $\delta \chi^{ijk} =  0$ can only be solved on the vacuum if
\begin{equation}
\label{A2=0}
{A_{2l}}^{ijk} =0.
\end{equation}
The vanishing of the gravitino variation leads instead to the killing spinor equation
\begin{equation}
\mathcal{D}_\mu \epsilon^i + \frac{\sqrt 2}{2}  A_1^{ij} \gamma_\mu \epsilon_j =0,
\end{equation}
which allows for a non-vanishing gravitino mass and thus for an Anti-de Sitter vacuum. 
Inserting \eqref{A2=0} into \eqref{Qm32}, we obtain 
\begin{equation}
\label{m32QN=8}
\mathcal{M}^{MN}{\mathcal{Q}_{Mi}}^j{\mathcal{Q}_{Nj}}^i = - 28 |{A_1}^{ij}|^2,
\end{equation}
which is the analogous of \eqref{ImNm32} and it enables us to express the gravitino mass in terms of the gauge charges and kinetic function. 
The vacuum energy becomes then
\begin{equation}
\label{ccN=8}
\mathcal{V}_{AdS} = -\frac 34  |A_1^{ij}|^2 = \frac{3}{112} \mathcal{M}^{MN} {Q_{Mi}}^j {Q_{Mj}^i} . 
\end{equation}

Without loss of generality and to better compare with the $N=2$ case, we can move to a purely electric frame. 
Reducing the symplectic indices $M, N,\dots$ to the electric ones $\Lambda,\Sigma,\dots$, we can rewrite \eqref{m32QN=8} as
\begin{equation}
- 28 |{A_1}^{ij}|^2= \mathcal{M}^{\Lambda \Sigma}{\mathcal{Q}_{\Lambda i}}^j{\mathcal{Q}_{\Sigma j}}^i = -(\rm{Im}\mathcal{N}^{-1})^{\Lambda \Sigma}{\mathcal{Q}_{\Lambda i}}^j{\mathcal{Q}_{\Sigma j}}^i .
\end{equation}
In the last step we used $ ({\rm Im} \mathcal{N}^{-1})^{\Lambda \Sigma}  = -\mathcal{M}^{\Lambda \Sigma}$, which can be derived from \eqref{ImN=LL}, \eqref{M=LL} and the coset relation ${L_{M}}^{ij} L_{N\,ij} - {L_{N}}^{ij} L_{M\,ij}  = i \Omega_{MN} $. 
In the purely electric frame, the vacuum energy \eqref{ccN=8} becomes then
\begin{equation}
\label{VN8AdSTrQQ}
 \mathcal{V}_{AdS} =- \frac{3}{112} ({\rm Im}\mathcal{N}^{-1})^{\Lambda \Sigma} {\mathcal{Q}_{\Lambda i}}^j {\mathcal{Q}^i}_{\Sigma j} = \frac{3}{112}  ({\rm Im}\mathcal{N}^{-1})^{\Lambda \Sigma} {\rm Tr} \mathcal{Q}^\dagger_{\Lambda} \mathcal{Q}_\Sigma.
\end{equation}
This is the analogous of formula \eqref{VN2AdSTrQQ} of the $N=2$ setup.

To recast the expression of the vacuum energy above in the form \eqref{Hquant}, we have now to identify the combination of vectors gauging the abelian factor associated to the weak gravity conjecture and canonically normalise the kinetic term. 
Given that we have already presented the manifestly covariant procedure in section \ref{N=2arg}, in the following we arrive at the same result by choosing a preferred direction among $\Lambda=0,\dots,27$ and thus avoid some level of complication. 
We follow closely the steps of \cite{DallAgata:2021nnr} (which were performed for the de Sitter case).

To canonically normalise the vectors, we introduce the vielbeins $\Theta^A_\Lambda$, such that
\begin{align}
- {\rm Im }\mathcal{N}_{\Lambda \Sigma} &= g^{-2} \, \delta_{AB} \Theta^A_{\Lambda} \Theta^B_\Sigma,\\ -\left({\rm Im}\mathcal{N}^{-1}\right)^{\Lambda \Sigma} &= g^2 \delta^{AB}\Theta_A^\Lambda \Theta_B^\Sigma,\\
 \Theta^A_\Lambda \Theta^\Lambda_B &= \delta^A_B, \qquad A, B= 1,\dots, 28.
\end{align}
In these expressions, $g$ is the abelian gauge coupling, which is in general function of the scalars
\begin{equation}
g = \sqrt{-\frac{1}{28} ({\rm Im}\mathcal{N}^{-1})^{\Lambda\Sigma}\, \delta_{AB}\Theta_\Lambda^A\Theta_\Sigma^B}\, ,
\end{equation}
analogous to \eqref{g32U12}. 
The canonically normalised vectors are then given by
\begin{equation}
v_\mu^A = \Theta^A_\Lambda A_\mu^\Lambda
\end{equation}
and the now hermitean charge matrix is
\begin{equation}
q_A = \frac i2 \Theta^\Lambda_A \mathcal{Q}_\Lambda,
\end{equation}
such that the gravitino covariant derivative reduces to
\begin{equation}
\mathcal{D}_{\mu} \psi_{\nu\, i} = \dots +i {v_\mu}^A q_{A\,i}^j\psi_{\nu j}.
\end{equation}
Finally, the vacuum energy \eqref{VN8AdSTrQQ} becomes
\begin{equation}
 \mathcal{V}_{AdS} = \frac{3}{112}({\rm Im}\mathcal{N}^{-1})^{\Lambda \Sigma} {\rm Tr} \mathcal{Q}_\Lambda^\dagger \mathcal{Q}_\Sigma = -\frac{3}{28} g^2 \delta^{AB} {\rm Tr}\, q_A q_B.
\end{equation}
Choosing for definiteness the direction of the weak gravity abelian group to be $A=1$, we can write ($q_1 \equiv q)$
\begin{equation}
 \mathcal{V}_{AdS} = -\frac{3}{28} g^2 \delta^{AB} {\rm Tr}\, q_A q_B \leq -\frac{3}{28} g^2  {\rm Tr}\, (q^2) , 
\end{equation}
which in absolute value means
\begin{equation}
|\mathcal{V}_{AdS}| \geq \frac{3}{28} g^2  {\rm Tr}\, (q^2) \gtrsim {\rm Tr}\, (q^2) \Lambda^2_{UV},
\end{equation}
where in the last step we enforced the magnetic weak gravity conjecture. 
As for the $N=2$ case, these vacua are precisely of the form \eqref{Hquant} and therefore cannot be scale separated.

Also in this case we should stress that while our argument assumes the existence of an abelian factor on the vacuum, it can be easily extended to non-abelian gauge groups along the lines discussed at the end of section \ref{sec:_n_2_supergravity}.
This is especially important in this case because we know that fully supersymmetric vacua of $N=8$ supergravity with a negative cosmological constant preserve an SO(8) gauge group \cite{DallAgata:2012mfj}.

\section{Partially supersymmetric vacua and possible loopholes}
\label{sec:partial_susy}

So far, we focussed on vacua preserving all supersymmetries. One could therefore ask if we can employ a similar argument for vacua where supersymmetry is partially broken.

A crucial aspect of our derivation is the fact that the theory under scrutiny should have a gauged U(1) symmetry under which the gravitini are charged.
We argue that if the (partially supersymmetry breaking) vacuum maintains this property, our argument still applies.
In fact, if a vacuum of a given $N'$-extended supergravity preserves $N<N'$ supersymmetries, one can construct a consistent truncation of the original theory to an $N$-extended supergravity along the lines of \cite{Andrianopoli:2001zh}.
In such a truncated model, the original vacuum appears as a fully supersymmetric one and the argument of the previous sections (generalised possibly to $2<N<8$) tells us once more that there is no scale separation.

Before discussing possible loopholes of this rather general argument, we would like to give an illustrative example of such a truncation.
For the sake of simplicity we focus on $N=2$ vacua of $N=8$ supergravity.
The reasons are mainly twofold: the maximal theory has a definite spectrum and hence one can fully classify its truncations; $N=2$ vacua fall into the proof given in section \ref{sec:_n_2_supergravity}.
While a general list of all possible $N=2$ truncations that are also resulting from spontaneous supersymmetry breaking is presented in the appendix, here we explicitly show how this reduction works for the case 2 of table~\ref{case14}, which we now discuss in some detail.
This case describes a model where the $N=8$ spectrum arranges at the $N=2$ vacuum in 15 vector multiplets, 10 hypermultiplets and 6 semilong gravitino multiplets in a way that allow for a consistent truncation to an $N=2$ supergravity with only 15 vector multiplets remaining, with scalar manifold
\begin{equation}
	{\cal M}_{scal} = \frac{{\rm SO}^*(12)}{{\rm U}(6)}.
\end{equation} 

To begin, let us state the transformation rules of the maximal theory at the 2-fermion level:
\begin{eqnarray}
  \label{eq:susy-transformations-gauged}
  \delta \psi_{\mu}{}^{i} 
  &=&
  2\,\mathcal{D}_{\mu}\epsilon^{i}
  +\frac{1}{4}\sqrt{2}\,{F}^{-}_{\rho\sigma}{}^{ij}\,
  \gamma^{\rho\sigma}\gamma_{\mu}\epsilon_{j}  +{\sqrt{2}}\,g \,A_1{}^{ij}\,\gamma_{\mu}\,\epsilon_{j} ,
     \nonumber\\[1ex]
  \delta \chi^{ijk} 
  &=&
  -2\sqrt{2}\,{\cal P}_{\mu}^{ijkl}\,\gamma^{\mu}\epsilon_{l}
  + \frac3{2} \, {F}^{-}_{\mu\nu}{}^{[ij}
  \gamma^{\mu\nu}\epsilon^{k]}  - 2 g\,A_{2 l}{}^{ijk}\,\epsilon^{l}   ,   \nonumber\\[1ex]
  \delta e_{\mu}{}^{a}&=&
  \bar\epsilon^{i}\gamma^{a}\psi_{\mu i} ~+~
  \bar\epsilon_{i}\gamma^{a}\psi_{\mu}{}^i , \nonumber\\[1ex]
  \delta{\cal V}_M{}^{ij} &=&  2\sqrt{2}\,{\cal V}_{M kl} \, \left(
  \bar\epsilon^{[i}\chi^{jkl]}+\frac1{24}\varepsilon^{ijklmnpq}\, 
  \bar\epsilon_{m}\chi_{npq}\right)   \,,  \nonumber \\[1ex]
    \delta A_{\mu}{}^{M}
    &=&
    - \mathrm{i}\,\Omega^{MN} {\cal V}_N{}^{ij}\,\left( 
    \bar\epsilon^{k}\,\gamma_{\mu}\,\chi_{ijk}
    +2\sqrt{2}\, \bar\epsilon_{i}\,\psi_{\mu j}\right)~+~ {\rm h.c.}\,,
\end{eqnarray}
where we recall that the covariant derivative of the supersymmetry parameters contain also the SU(8) composite connection, $Q_\mu{}^i{}_j$, i.e.~
\begin{equation}
		{\cal D}_\mu \epsilon^i = \partial_\mu\epsilon^i - \frac14 \, \omega_\mu{}^{ab} \gamma_{ab} \epsilon^i + \frac12\, Q_\mu{}^i{}_j \epsilon^j.
\end{equation}
It is also useful to rewrite the scalar transformation as 
\begin{equation}
		\delta \phi^u P_u^{ijkl} = 2 \sqrt2\, \left(
  \bar\epsilon^{[i}\chi^{jkl]}+\frac1{24}\varepsilon^{ijklmnpq}\, 
  \bar\epsilon_{m}\chi_{npq}\right) ,
\end{equation}
where we the employed the coset manifold properties of the scalar manifold.

If we have a vacuum preserving 2 supersymmetries, we can split the supersymmetry parameter accordingly:
\begin{equation}
		\epsilon^i = \{ \epsilon^A, \epsilon^{\underline{a}}\},
\end{equation}
where $A = 1,2$ specifies the preserved transformations (rotating under the residual U(2) R-symmetry) and $\underline{a} = 1,\ldots,6$ counts the supersymmetries that are violated on the vacuum, which we assume for now to be transforming in the fundamental representation of SU(6).
The truncation of the theory to the fields and transformations preserving the 2 residual supersymmetries described by $\epsilon^A$ can be obtained by following the prescription in \cite{Andrianopoli:2001zh}.
This effectively amounts to set $\epsilon^{\underline{a}} = 0$ consistently in the transformation rules above.
Also, since we want an ordinary gauged $N=2$ supergravity as a result, we should truncate away the corresponding gravitini, $\psi_\mu{}^{\underline{a}} = 0$.

Once we discard the $\psi_\mu{}^{\underline{a}}$ gravitini, we need to prove that its supersymmetry transformation under the residual supersymmetries consistently vanishes, i.e.~
\begin{equation}
		\delta \psi_\mu{}^{\underline{a}} = 0 = Q_{\mu \underline{A}}{}^a \epsilon^A + \frac{1}{4}\sqrt{2}\,{F}^{-}_{\rho\sigma}{}^{\underline{a} A}\,
  \gamma^{\rho\sigma}\gamma_{\mu}\epsilon_{A}  +{\sqrt{2}}\,g \,A_1{}^{\underline{a} A}\,\gamma_{\mu}\,\epsilon_{A}.
\end{equation}
This requires that we truncate
\begin{equation}
		 Q_{\mu {A}}{}^{\underline{a}} = 0 = F_{\rho\sigma}{}^{\underline{a} A}  = A_1{}^{\underline{a} A}.
\end{equation}
The first condition implies the factorisation of the scalar manifold.
In detail, we can check the components of the scalar curvature and see that 
\begin{equation}\label{factores}
		d Q_A{}^{\underline{a}} + Q_A{}^i \wedge Q_i^{\underline{a}} = 0 = P^{\underline{a}ijk} \wedge P_{A ijk} = 3 P^{\underline{acd}B} \wedge P_{AB \underline{cd}} + P^{\underline{abcd}} \wedge P_{A \underline{bcd}}.
\end{equation}
This in turn implies that either
\begin{equation}\label{scalar1}
		P_{AB \underline{ab}} = 0 = P_{\underline{abcd}},
\end{equation}
or
\begin{equation}\label{scalar2}
		P_{A \underline{abc}} = 0,
\end{equation}
as one can see from (\ref{factores}) by employing the duality properties of the vielbein, which tells us that \begin{equation}
		P^{\underline{adef}} \wedge P_{A \underline{def}}  = \frac{1}{4}\epsilon^{\underline{abcdef}}\, \epsilon^{CD} P_{CD \underline{bc}} \wedge P_{A \underline{def}} = 0,
\end{equation}
Let us analyse the two cases separately.

In the case we impose (\ref{scalar2}), consistency further requires that
\begin{equation}
		\delta z^i P_i^{A \underline{abc}} = \sqrt2\, \left(
  \bar\epsilon^{A}\chi^{\underline{abc}}\right) = 0,
\end{equation}
which means that we have to remove from the spectrum 20 fermions
\begin{equation}
		\chi^{\underline{abc}} = 0.
\end{equation}
Moreover, having removed the vector combinations $F^{A \underline{a}} \equiv F^M {\cal V}_M{}^{A \underline{a}}$, consistency of the transformation rules of the corresponding vector field combinations $\delta A^M {\cal V}_M{}^{A \underline{a}} = 0$ forces to remove other 6 fermions
\begin{equation}
		\chi^{AB \underline{a}} = 0
\end{equation}
and the fermion shifts
\begin{equation}
		A_C{}^{AB \underline{a}} = 0.
\end{equation}
Finally, consistency of the fermion supersymmetry rules further constraints the fermion shifts:
\begin{equation}
		\delta \chi^{\underline{abc}} = 0 =   A_C{}^{\underline{abc}}.
\end{equation}
Altogether this results in the following residual supersymmetry variations:
\begin{eqnarray}
  \delta \psi_{\mu}{}^{A} 
  &=&
  2\,\mathcal{D}_{\mu}\epsilon^{A}
  +\frac{1}{4}\sqrt{2}\,{F}^{-}_{\rho\sigma}{}^{AB}\,
  \gamma^{\rho\sigma}\gamma_{\mu}\epsilon_{B}  +{\sqrt{2}}\,g \,A_1{}^{AB}\,\gamma_{\mu}\,\epsilon_{B} ,
     \nonumber\\[1ex]
  \delta \chi^{A\underline{ab}} 
  &=&
  -2\sqrt{2}\,{\cal P}_{\mu}^{\underline{ab}AB}\,\gamma^{\mu}\epsilon_{B} - 2 g\,A_{2 B}{}^{A\underline{ab}}\,\epsilon^{B}   ,   \nonumber\\[1ex]
  \delta e_{\mu}{}^{a}&=&
  \bar\epsilon^{A}\gamma^{a}\psi_{\mu A} ~+~
  \bar\epsilon_{A}\gamma^{a}\psi_{\mu}{}^A , \nonumber\\[1ex]
		\delta z^i P_i^{\underline{ab}AB} &=& \frac{1}{\sqrt2}\, \left(
  \bar\epsilon^{[A}\chi^{B]\underline{ab}}\right), \\[1ex]
    \delta A_{\mu}{}^{M}
    &=&
    - \mathrm{i}\,\Omega^{MN} {\cal V}_N{}^{AB}\,\left( 
	2\sqrt{2}\, \bar\epsilon_{A}\,\psi_{\mu B}\right) \,
    - \mathrm{i}\,\Omega^{MN} {\cal V}_N{}^{\underline{ab}}\,\left( 
    \bar\epsilon^{A}\,\gamma_{\mu}\,\chi_{\underline{ab}A}\right)~+~ {\rm h.c.}\,.\nonumber
\end{eqnarray}
The fields that survive clearly form a gravity multiplet and 15 vector multiplets.
There are indeed fermions in the $(\mathbf{15},\mathbf{2})$ of SU(6) $\times$ SU(2), scalars in the $(\mathbf{15},\mathbf{1})$ and $(\overline{\mathbf{15}},\mathbf{1})$ and the vectors surviving are also in the $(\mathbf{15},\mathbf{1}) \oplus (\mathbf{1},\mathbf{1})$.
Moreover, the fields we truncated away are also filling proper supersymmetric representations with respect to $N=2$ supersymmetry, if we assume that the supersymmetry breaking vacuum has vanishing cosmological constant.
In this case, we find that the 6 gravitini arrange themselves together with the other fields in 6 semilong gravitino multiplets and the remaining fields can be collected in 10 hypermultiplets.

In the case we decided to impose the other condition, namely (\ref{scalar1}), the same consistency relations would have required setting
\begin{equation}
\begin{split}
	P_{AB \underline{ab}} = P_{\underline{abcd}} = 0, \\[2mm]
	\chi^{A \underline{ab}} = \chi^{AB \underline{a}} = F^{A \underline{a}} = 
	    \chi_{BA \underline{a}} =0, \\[2mm]
		A_C{}^{AB \underline{a}} = F^{\underline{ab}} =  A_C{}^{A \underline{ab}} =0 .
\end{split}		
\end{equation}
This results in a truncated theory with the gravity multiplet coupled to fermions in the $\mathbf{20}$ of SU(6) and fermions in the $(\mathbf{20},\mathbf{2})$ of SU(6) $\times$ SU(2), which can be interpreted as the coupling of 20 hypermultiplets.
The truncated fields though cannot be arranged in a consistent way in semilong or long representations of $N=2$ supersymmetry, neither in a Minkowski nor in an Anti-de Sitter vacuum. 
This therefore means that we cannot obtain such a model from a spontaneous supersymmetry breaking scenario.

We can now comment on possible loopholes of our general argument for vacua preserving $N$ supersymmetries in $N'$-extended supergravity theories.
The first one is that the abelian factor used in the original $N'>N$ extended supergravity may not survive the truncation.
This is certainly an element of concern, but since the $N$ supersymmetric vacuum is still a vacuum whose cosmological constant is generated by a gauging procedure (assuming $N>1$), its value must be related to some charges of the residual gauge group of the truncated theory and we should still be able to apply our argument, at least in the weaker form for non-abelian factors.

Another reason of concern is the use of the KK scale as a proxy for the ultraviolet cutoff, $\Lambda_{KK} \sim \Lambda_{UV}$. This point can be subtle as a correct estimation of the KK scale might be highly non-trivial, depending on the precise details of the compact manifold. We refer to \cite{DeLuca:2021mcj} for recent work in this direction, in particular taking into account warping effects. If the ultraviolet cutoff turns out to be bigger than the actual KK scale, the effective description would break down anyway. However, if the cutoff is smaller the theory might be scale separated. Indeed, one could imagine the presence of another tower of states which are lighter than KK states. At $\Lambda_{UV}\sim \Lambda_{tower}$ the effective description breaks down, but \eqref{Hquant} is not giving us information on scale separation. In other words, our argument can be interpreted as the statement that supersymmetric anti-de Sitter vacua cannot be scale separated if the supergravity theory is valid up to the KK scale. Notice that this loophole is in tension with the emergent string conjecture \cite{Lee:2019wij}, which suggests that the existence of a (string) scale parametrically smaller than $\Lambda_{KK}$ is pathological and indeed never realised in any known example \cite{Klaewer:2020lfg}.

Finally, yet another loophole is the possibility that the vacuum is generated in an $N=1$ theory in a way that it cannot be related to any consistent truncation of an extended supergravity. 
This can happen if the model contains sources that break supersymmetry and at the same time violate the Jacobi identities of any candidate $N>1$ parent theory.
The vacuum should be $N=1$, otherwise our argument, which crucially uses gauged supergravity, would still apply.
One could in fact build $N=1$ models where the superpotential and D-terms cannot be expressed as truncations of an $N>1$ theory. 
An example of how this can happen can be seen in \cite{Villadoro:2005cu}, where $N=1$ IIA vacua and effective actions were discussed.
The origin of the $N=1$ theory comes from the superposition of orbifolds and orientifolds that preserve different sets of the original supersymmetries.
This means that while each system of branes and orientifold can preserve half of the original supersymmetry and be described by a $N=4$ gauged supergravity, whose Jacobi identities correspond to the fluxes and localised sources, the simultaneous presence of different projections renders the final theory genuinely $N=1$.
Actually, the precise relation between orbifolds respecting the complex structure of the underlying manifold and gauged supergravities has been worked out in \cite{Condeescu:2013yma}.
We should also stress that the presence of orientifolds as a necessary ingredient is somehow expected to obtain scale separated vacua, because they allow the violation of generic energy conditions respected by other sources.

\section{Conclusion}
\label{sec:conclusion}

In this work, we investigated the existence of scale separated Anti-de Sitter vacua of extended supergravity in four dimensions. We presented a general argument showing that such vacua are in tension with the magnetic gravity conjecture. For $N=2$ and $N=8$ vacua our analysis was explicit, and we explained how it can be extended to vacua with partial breaking of supersymmetry fitting in the between. A necessary working assumption is the presence of an unbroken abelian factor in the vacuum, needed to consistently apply the weak gravity conjecture. In case only non-abelian groups survive, we can still give a weaker version of our argument, along the lines of \cite{DallAgata:2021nnr}. 
While one could envisage vacua where the residual gauge symmetry group is empty (see for instance \cite{Fischbacher:2009cj} for maximal supergravity), these are clearly uninteresting for phenomenological applications and in any case they mostly arise from models which are ruled out by the presence of other vacua with residual abelian factors.

Our results indicate that, contrary to what one might expect, scale separation is not straightforward to obtain already at the effective field theory level, if the amount of preserved supersymmetry is not minimal or even absent. Indeed, the most realistic way we foresee to circumvent our argument is to consider (at most) $N=1$ theories in four dimensions which cannot be obtained as consistent truncations of a parent theory with a higher amount of preserved supersymmetry.\footnote{The analogy with the fate of de Sitter vacua seems to hold also in this respect, as \cite{Andriot:2022way} recently conjectured that they should be possible only within (at most) $N=1$ supersymmetric theories in four dimensions.} This could explain why certain classes of string compactifications with residual $N=0,1$ supersymmetry in four dimensions, such as \cite{DeWolfe:2005uu}, seem indeed to be scale separated. 

The work here presented can be extended along various directions. For example, our analysis in section \ref{sec:partial_susy} was meant to give a proof of principle that the argument should work more generally in any $N$-extended supergravity. It would be of interest, even in its own rights, to perform a detailed analysis of the possible patterns of partial supersymmetry breaking of the maximal theory and construct explicitly the corresponding actions. With respect to the scale separation problem, a major step forward would clearly be to understand precisely if and how $N=0,1$ Anti-de Sitter vacua do indeed evade our argument in the way we suggested above, or perhaps in some other manner. It would be also interesting to extend the argument to spacetime dimensions higher than four. Indeed, in $d>4$ dimensions the scalar potential of gauged supergravity necessarily stems from a gauging procedure and thus we believe that our argument applies to such models as well. If true, this could exclude scale separation for anti-de Sitter vacua in $d>4$, $N>0$ theories. We leave an explicit verification of this statement for future work.

\section*{Acknowledgements}

We acknowledge discussions on the topic of this work with F.~Farakos, E.~Palti, T.~van Riet and especially L.~Martucci, who we thank for suggesting first to investigate the connection between our previous work and the scale separation problem. 
The work of N.C.~is supported by the Alexander von Humboldt foundation.
The work of G.D.~is partially supported by the MIUR-PRIN contract 2017CC72MK003.

\section*{Appendix: Supersymmetry breaking $N=8$ to $N=2$}

For completeness, we report in this appendix tables with the data regarding the possible arrangements of the $N=8$ supergravity degrees of freedom into $N=2$ multiplets in both Minkowski and de Sitter spacetime, therefore allowing for a spontaneous breaking of supersymmetry.

The degrees of freedom of the $N=8$ supergravity multiplet and those of the $N=2$ massless and massive multiplets are listed in table~\ref{table1}.

\renewcommand{\arraystretch}{1.3}

\begin{table}[h!]
	\begin{center}
\begin{tabular}{ccccccc}\hline\hline
& multiplet & spin 2 & spin 3/2  & spin 1  & spin 1/2  &  spin 0 \\\hline
& $N=8$ gravity & 1 & 8 & 28 & 56 & 70 \\\hline
& $N=2$ gravity & 1 & 2 & 1 & 0 & 0 \\
$\psi_l$ & $N=2$ long gravitino & 0 & 1 & 4 & 7 & 8 \\
$\psi_s$ & $N=2$ semilong Mink. gravitino ($2\times$) & 0 & 1 & 2 & 1 & 0 \\
$\psi_s$ & $N=2$ semilong AdS gravitino ($2\times$) & 0 & 1 & 3 & 4 & 4 \\
$V$ & $N=2$ short vector & 0 & 0 & 1 & 2 & 2 \\
$v_l$ & $N=2$ long vector & 0 & 0 & 1 & 4 & 6 \\
$v_s$ & $N=2$ semilong Mink. vector ($2\times$) & 0 & 0 & 1 & 2 & 2 \\
$v_s$ & $N=2$ semilong AdS vector ($2\times$) & 0 & 0 & 1 & 3 & 4 \\
$H$ & $N=2$ hyper & 0 & 0 & 0 & 2 & 4 \\\hline\hline
	\end{tabular}
	\end{center}
	\caption{Multiplet content. We listed the states by helicities. For massive multiplets, each massive vector multiplets will eat one of the scalar degrees of freedom and each gravitino will eat one of the spin 1/2 degrees of freedom.}
	\label{table1}
\end{table}

We also provide tables for each of the 11 possible routes of partial supersymmetry breaking, giving the symmetry breaking pattern, the number of surviving hyper $n_H$ and short vector $n_V$ multiplets and the spectrum of the truncated states (as noted in \cite{Andrianopoli:2001zh}, additional cases arise as further truncations of these ``maximal'' cases).
These cases follow from the classification of the possible truncations in \cite{Andrianopoli:2001zh}.
The novelty here is that we analysed if and how the matter fields in such truncations could be arranged in consistent representations of $N=2$ supersymmetry, so that we could interpret them as the result of a spontaneous supersymmetry breaking scenario.

In detail, the 11 truncation patterns and their relation to the residual scalar manifolds are summarised in tables \ref{scalar}-\ref{case11}.

\begin{table}[h!]
	\begin{center}
	\begin{tabular}{cc}\hline\hline
		Case & scalar manifold \\\hline
		\\
		1 & $\frac{{\rm E}_{6(2)}}{{\rm SU}(6) \times {\rm SU}(2)}$ \\[4mm]
		2 & $\frac{{\rm SO}^*(12)}{{\rm U}(6)}$ \\[4mm]
		3 & $\frac{{\rm SU}(3,3)}{{\rm SU}(3) \times {\rm SU}(3) \times {\rm U}(1)} \times \frac{{\rm SU}(2,1)}{{\rm SU}(2) \times {\rm U}(1)}$ \\[4mm]
		4 & $\frac{{\rm Sp}(6)}{{\rm U}(3)} \times \frac{{\rm G}_{2(2)}}{{\rm SO}(4)}$ \\[4mm]
		5 & $\frac{{\rm SU}(1,1)}{{\rm U}(1)} \times \frac{{\rm F}_{4(4)}}{{\rm USp}(6) \times {\rm USp}(2)}$ \\[4mm]
		6 & $\frac{{\rm SU}(1,1)}{{\rm U}(1)} \times \frac{{\rm SO}(6,4)}{{\rm SO}(6) \times {\rm SO}(4)}$ \\[4mm]
		7 & $\left[\frac{{\rm SU}(1,1)}{{\rm U}(1)}\right]^2 \times \frac{{\rm SO}(4,5)}{{\rm SO}(4) \times {\rm SO}(5)}$ \\[4mm]
		8 & $\left[\frac{{\rm SU}(1,1)}{{\rm U}(1)}\right]^3 \times \frac{{\rm SO}(4,4)}{{\rm SO}(4) \times {\rm SO}(4)}$ \\[4mm]
		9 & $\left[\frac{{\rm SU}(1,1)}{{\rm U}(1)} \times \frac{{\rm SO}(2,3)}{{\rm SO}(2) \times {\rm SO}(3)}\right] \times \frac{{\rm SO}(4,3)}{{\rm SO}(4) \times {\rm SO}(3)}$ \\[4mm]
		10 & $\left[\frac{{\rm SU}(1,1)}{{\rm U}(1)} \times \frac{{\rm SO}(2,4)}{{\rm SO}(2) \times {\rm SO}(4)}\right] \times \frac{{\rm SO}(4,2)}{{\rm SO}(4) \times {\rm SO}(2)}$ \\[4mm]
		11 & $\left[\frac{{\rm SU}(1,1)}{{\rm U}(1)} \times \frac{{\rm SO}(2,5)}{{\rm SO}(2) \times {\rm SO}(5)}\right] \times \frac{{\rm SO}(4,1)}{{\rm SO}(4)}$ \\[4mm]
		\hline\hline
	\end{tabular}
	\end{center}
	\caption{Scalar manifolds of the truncated theories}
	\label{scalar}
\end{table}

\begin{table}[h!]
	\begin{center}
	\begin{tabular}{ccccccccc}\hline\hline
		Case & c.c.  & \multicolumn{2}{c}{surviving spectrum} & \multicolumn{5}{c}{truncated spectrum} \\
		&& $n_V$ & $n_H$ & $n_{\psi_l}$ &  $n_{\psi_s}$ & $n_{v_l}$ & $n_{v_s}$ & $H$ \\\hline
		1  & Mink/AdS  & 0 &10 & \multicolumn{5}{c}{no consistent arrangement} \\
		2  & Mink & 15 &0& 0 & 6 & 0 & 0 & 10 \\
		2  & AdS & 15 &0& \multicolumn{5}{c}{no consistent arrangement} \\
		3 & Mink  & 9 & 1 & 2 & 4 & 2 & 0 & 5 \\
		3 & Mink  & 9 & 1 & 2 & 4 & 0 & 2 & 7 \\
		3 & Mink  & 9 & 1 & 0 & 6 & 6 & 0 & 3 \\
		3 & Mink  & 9 & 1 & 0 & 6 & 4 & 2 & 5 \\
		3 & Mink  & 9 & 1 & 0 & 6 & 2 & 4 & 7 \\
		3 & Mink  & 9 & 1 & 0 & 6 & 0 & 6 & 9 \\
		3 & AdS  & 9 & 1 & 0 & 6 & 0 & 0 & 6 \\		
		\hline\hline
	\end{tabular}
	\end{center}
	\caption{Supersymmetry breaking patterns and truncation of the spectrum -- cases 1-3}
	\label{case14}
\end{table}

\begin{table}[h!]
	\begin{center}
	\begin{tabular}{ccccccccc}\hline\hline
		Case & c.c.  & \multicolumn{2}{c}{surviving spectrum} & \multicolumn{5}{c}{truncated spectrum} \\
		&& $n_V$ & $n_H$ & $n_{\psi_l}$ &  $n_{\psi_s}$ & $n_{v_l}$ & $n_{v_s}$ & $H$ \\\hline
		4  & Mink   & 6 & 2 & 4 & 2 & 1 & 0 & 3 \\
		4  & Mink   & 6 & 2 & 2 & 4 & 5 & 0 & 1 \\
		4  & Mink   & 6 & 2 & 2 & 4 & 3 & 2 & 3 \\		
		4  & Mink   & 6 & 2 & 2 & 4 & 1 & 4 & 5 \\
		4  & Mink   & 6 & 2 & 0 & 6 & 7 & 2 & 1 \\
		4  & Mink   & 6 & 2 & 0 & 6 & 5 & 4 & 3 \\
		4  & Mink   & 6 & 2 & 0 & 6 & 3 & 6 & 5 \\
		4  & Mink   & 6 & 2 & 0 & 6 & 1 & 8 & 7 \\
		4  & AdS  & 6 & 2 & 2 & 3 & 1 & 0 & 3 \\
		4  & AdS  & 6 & 2 & 0 & 6 & 3 & 0 & 2 \\
		4  & AdS  & 6 & 2 & 0 & 6 & 1 & 2 & 3 \\
		\hline\hline
	\end{tabular}
	\end{center}
	\caption{Supersymmetry breaking patterns and truncation of the spectrum -- case 4}
	\label{case4}
\end{table}

\begin{table}[h!]
	\begin{center}
	\begin{tabular}{ccccccccc}\hline\hline
		Case & c.c.  & \multicolumn{2}{c}{surviving spectrum} & \multicolumn{5}{c}{truncated spectrum} \\
		&& $n_V$ & $n_H$ & $n_{\psi_l}$ &  $n_{\psi_s}$ & $n_{v_l}$ & $n_{v_s}$ & $H$ \\\hline
		5  & AdS   & 1 & 7 & \multicolumn{5}{c}{no consistent arrangement} \\
		5  & Mink  & 1 & 7 & 2 & 4 & 0 & 10 & 1 \\
		5  & Mink  & 1 & 7 & 0 & 6 & 2 & 12 & 1 \\
		5  & Mink  & 1 & 7 & 0 & 6 & 0 & 14 & 3 \\
		6  & AdS  & 1 & 6 & \multicolumn{5}{c}{no consistent arrangement} \\
		6  & Mink  & 1 & 6 & 4 & 2 & 0 & 6 & 0 \\
		6  & Mink  & 1 & 6 & 2 & 4 & 2 & 8 & 0 \\
		6  & Mink  & 1 & 6 & 2 & 4 & 0 & 10 & 2 \\
		6  & Mink  & 1 & 6 & 0 & 6 & 4 & 10 & 0 \\
		6  & Mink  & 1 & 6 & 0 & 6 & 2 & 12 & 2 \\
		6  & Mink  & 1 & 6 & 0 & 6 & 0 & 14 & 4 \\
		7  & AdS  & 2 & 5 & \multicolumn{5}{c}{no consistent arrangement} \\
		7  & Mink  & 2 & 5 & 4 & 2 & 1 & 4 & 0 \\
		7  & Mink  & 2 & 5 & 2 & 4 & 3 & 6 & 0 \\
		7  & Mink  & 2 & 5 & 2 & 4 & 1 & 8 & 2 \\
		7  & Mink  & 2 & 5 & 0 & 6 & 5 & 8 & 0 \\
		7  & Mink  & 2 & 5 & 0 & 6 & 3 & 10 & 2 \\
		7  & Mink  & 2 & 5 & 0 & 6 & 1 & 12 & 4 \\
		\hline\hline
	\end{tabular}
	\end{center}
	\caption{Supersymmetry breaking patterns and truncation of the spectrum -- cases 5-7}
	\label{case56}
\end{table}

\begin{table}[h!]
	\begin{center}
	\begin{tabular}{ccccccccc}\hline\hline
		Case & c.c.  & \multicolumn{2}{c}{surviving spectrum} & \multicolumn{5}{c}{truncated spectrum} \\
		&& $n_V$ & $n_H$ & $n_{\psi_l}$ &  $n_{\psi_s}$ & $n_{v_l}$ & $n_{v_s}$ & $H$ \\\hline
		8  & AdS  & 3 & 4 & 6 & 0 & 0 & 0 & 0 \\
		8  & AdS  & 3 & 4 & 4 & 2 & 0 & 2 & 0 \\
		8  & AdS  & 3 & 4 & 2 & 4 & 0 & 4 & 0 \\
		8  & AdS  & 3 & 4 & 0 & 6 & 0 & 6 & 0 \\
		8  & Mink  & 3 & 4 & 6 & 0 & 0 & 0 & 0 \\
		8  & Mink  & 3 & 4 & 4 & 2 & 2 & 2 & 0 \\
		8  & Mink  & 3 & 4 & 4 & 2 & 0 & 4 & 2 \\
		8  & Mink  & 3 & 4 & 2 & 4 & 4 & 4 & 0 \\
		8  & Mink  & 3 & 4 & 2 & 4 & 2 & 6 & 2 \\
		8  & Mink  & 3 & 4 & 2 & 4 & 0 & 8 & 4 \\
		8  & Mink  & 3 & 4 & 0 & 6 & 6 & 6 & 0 \\
		8  & Mink  & 3 & 4 & 0 & 6 & 4 & 8 & 2 \\
		8  & Mink  & 3 & 4 & 0 & 6 & 2 & 10 & 4 \\
		8  & Mink  & 3 & 4 & 0 & 6 & 0 & 12 & 6 \\
		\hline\hline
	\end{tabular}
	\end{center}
	\caption{Supersymmetry breaking patterns and truncation of the spectrum -- case 8}
	\label{case8}
\end{table}

\begin{table}[h!]
	\begin{center}
	\begin{tabular}{ccccccccc}\hline\hline
		Case & c.c.  & \multicolumn{2}{c}{surviving spectrum} & \multicolumn{5}{c}{truncated spectrum} \\
		&& $n_V$ & $n_H$ & $n_{\psi_l}$ &  $n_{\psi_s}$ & $n_{v_l}$ & $n_{v_s}$ & $H$ \\\hline
		9  & AdS  & 4 & 3 & 4 & 2 & 1 & 0 & 1 \\
		9  & AdS  & 4 & 3 & 2 & 4 & 3 & 0 & 0 \\
		9  & AdS  & 4 & 3 & 2 & 4 & 1 & 2 & 1 \\
		9  & AdS  & 4 & 3 & 0 & 6 & 3 & 2 & 0 \\
		9  & AdS  & 4 & 3 & 0 & 6 & 1 & 4 & 1 \\
		9  & Mink  & 4 & 3 & 4 & 2 & 3 & 0 & 0 \\
		9  & Mink  & 4 & 3 & 4 & 2 & 1 & 2 & 2 \\
		9  & Mink  & 4 & 3 & 2 & 4 & 5 & 2 & 0 \\
		9  & Mink  & 4 & 3 & 2 & 4 & 3 & 4 & 2 \\
		9  & Mink  & 4 & 3 & 2 & 4 & 1 & 6 & 4 \\
		9  & Mink  & 4 & 3 & 0 & 6 & 7 & 4 & 0 \\
		9  & Mink  & 4 & 3 & 0 & 6 & 5 & 6 & 2 \\
		9  & Mink  & 4 & 3 & 0 & 6 & 3 & 8 & 4 \\
		9  & Mink  & 4 & 3 & 0 & 6 & 1 & 10 & 6 \\
		\hline\hline
	\end{tabular}
	\end{center}
	\caption{Supersymmetry breaking patterns and truncation of the spectrum -- case 9}
	\label{case9}
\end{table}

\begin{table}[h!]
	\begin{center}
	\begin{tabular}{ccccccccc}\hline\hline
		Case & c.c.  & \multicolumn{2}{c}{surviving spectrum} & \multicolumn{5}{c}{truncated spectrum} \\
		&& $n_V$ & $n_H$ & $n_{\psi_l}$ &  $n_{\psi_s}$ & $n_{v_l}$ & $n_{v_s}$ & $H$ \\\hline
		10  & AdS  & 5 & 2 & 4 & 2 & 0 & 0 & 3 \\
		10  & AdS  & 5 & 2 & 2 & 4 & 2 & 0 & 2 \\
		10  & AdS  & 5 & 2 & 2 & 4 & 0 & 2 & 3 \\
		10  & AdS  & 5 & 2 & 0 & 6 & 4 & 0 & 1 \\
		10  & AdS  & 5 & 2 & 0 & 6 & 2 & 2 & 2 \\
		10  & AdS  & 5 & 2 & 0 & 6 & 0 & 4 & 3 \\
		10  & Mink  & 5 & 2 & 4 & 2 & 2 & 0 & 2 \\
		10  & Mink  & 5 & 2 & 4 & 2 & 0 & 2 & 4 \\
		10  & Mink  & 5 & 2 & 2 & 4 & 6 & 0 & 0 \\
		10  & Mink  & 5 & 2 & 2 & 4 & 4 & 2 & 2 \\
		10  & Mink  & 5 & 2 & 2 & 4 & 2 & 4 & 4 \\
		10  & Mink  & 5 & 2 & 0 & 6 & 0 & 6 & 6 \\
		10  & Mink  & 5 & 2 & 0 & 6 & 8 & 2 & 0 \\
		10  & Mink  & 5 & 2 & 0 & 6 & 6 & 4 & 2 \\
		10  & Mink  & 5 & 2 & 0 & 6 & 4 & 6 & 4 \\
		10  & Mink  & 5 & 2 & 0 & 6 & 2 & 8 & 6 \\
		10  & Mink  & 5 & 2 & 0 & 6 & 0 & 10 & 8 \\
		\hline\hline
	\end{tabular}
	\end{center}
	\caption{Supersymmetry breaking patterns and truncation of the spectrum -- case 10}
	\label{case10}
\end{table}

\begin{table}[h!]
	\begin{center}
	\begin{tabular}{ccccccccc}\hline\hline
		Case & c.c.  & \multicolumn{2}{c}{surviving spectrum} & \multicolumn{5}{c}{truncated spectrum} \\
		&& $n_V$ & $n_H$ & $n_{\psi_l}$ &  $n_{\psi_s}$ & $n_{v_l}$ & $n_{v_s}$ & $H$ \\\hline
		11  & AdS  & 6 & 1 & 2 & 4 & 0 & 0 & 5 \\
		11  & AdS  & 6 & 1 & 0 & 6 & 2 & 0 & 4 \\
		11  & AdS  & 6 & 1 & 0 & 6 & 0 & 2 & 5 \\
		11  & Mink  & 6 & 1 & 4 & 2 & 0 & 0 & 5 \\
		11  & Mink  & 6 & 1 & 4 & 4 & 4 & 0 & 3 \\
		11  & Mink  & 6 & 1 & 2 & 4 & 2 & 2 & 5 \\
		11  & Mink  & 6 & 1 & 2 & 4 & 0 & 4 & 7 \\
		11  & Mink  & 6 & 1 & 2 & 6 & 8 & 0 & 1 \\
		11  & Mink  & 6 & 1 & 0 & 6 & 6 & 2 & 3 \\
		11  & Mink  & 6 & 1 & 0 & 6 & 4 & 4 & 5 \\
		11  & Mink  & 6 & 1 & 0 & 6 & 2 & 6 & 7 \\
		11  & Mink  & 6 & 1 & 0 & 6 & 0 & 8 & 9 \\
		\hline\hline
	\end{tabular}
	\end{center}
	\caption{Supersymmetry breaking patterns and truncation of the spectrum -- case 11}
	\label{case11}
\end{table}

\end{document}